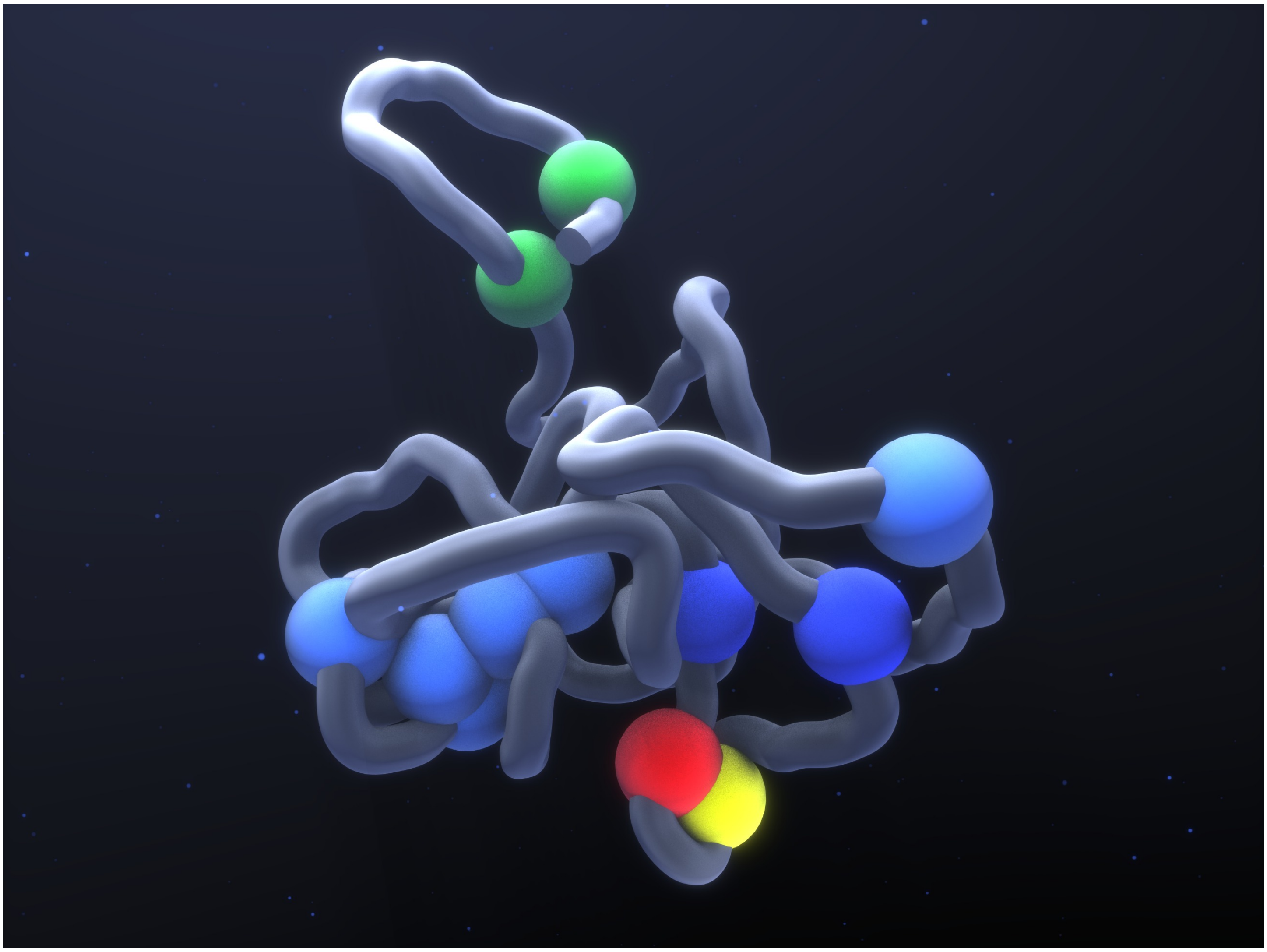

Classification: Computational Biology; Gene Regulation, Chromatin and Epigenetics

# CTCF-mediated transcriptional regulation through cell type-specific chromosome organization in the β-globin locus

Ivan Junier[1,2,3*], Ryan Dale[4], Chunhui Hou[4#], François Képès[1*] and Ann Dean[4*]

[1]Epigenomics Project and institute of Systems and Synthetic Biology, Genopole®, CNRS, University of Evry, Evry, France

[2]Institute of Complex Systems, Paris, France

[3]Centre for Genomic Regulation, Barcelona, Spain

[4]Laboratory of Cellular and Developmental Biology, National Institute of Diabetes and Digestive and Kidney Diseases, National Institutes of Health, Bethesda, MD 20892, USA.

*To whom correspondence should be addressed:
Ann Dean
Tel:1-301-496-6068; FAX: 1-301-496-6068; Email: anndean@helix.nih.gov
Laboratory of Cellular and Developmental Biology, NIDDK, NIH
50 South Drive, MSC 8028, Bethesda, MD 20892 USA

Correspondence may also be addressed to:
Francois Kepes
Tel: +33 169474431; FAX: +33 169474437;
Email: francois.kepes@epigenomique.genopole.fr
Epigenomics Project, Genopole Campus 1-Genavenir 6
5 rue Henri Desbrueres, Evry, France F-91030

Ivan Junier
Tel: +34-93-3160166; FAX: +34-93-3969983; Email: i.junier@gmail.com
Centre for Genomic Regulation, C/ dr. Aiguader, 88, 08003 Barcelona, Spain

[#]Current address: Department of Biology, Emory University, Atlanta, GA, USA

Key words: CTCF, β-globin genes, chromatin loops, polymer physics

**ABSTRACT**

The principles underlying the architectural landscape of chromatin beyond the nucleosome level in living cells remains largely unknown despite its potential to play a role in mammalian gene regulation. We investigated the 3-dimensional folding of a 1 Mbp region of human chromosome 11 containing the β-globin genes by integrating looping interactions of the insulator protein CTCF determined comprehensively by chromosome conformation capture (3C) into a polymer model of chromatin. We find that CTCF-mediated cell type specific interactions in erythroid cells are organized to favor contacts known to occur *in vivo* between the β-globin locus control region (LCR) and genes. In these cells, the modeled β-globin domain folds into a globule with the LCR and the active globin genes on the periphery. By contrast, in non-erythroid cells, the globule is less compact with few but dominant CTCF interactions driving the genes away from the LCR. This leads to a decrease in contact frequencies that can exceed 1000-fold depending on the stiffness of the chromatin and the exact positioning of the genes. Our findings show that an ensemble of CTCF contacts functionally affects spatial distances between control elements and target genes contributing to chromosomal organization required for transcription.

## INTRODUCTION

Chromatin interactions that form loops between enhancers and target promoters over large linear distances are common in metazoans (1-3). The specificity of such interactions is thought to be influenced by DNA-protein complexes called insulators, also through the formation of looped domains (4). How enhancer-promoter proximity is established and how topological domains contribute to transcription activation is not known. Moreover, recent technological advances have revealed genome-wide networks of intra- and inter-chromosomal sites of contact (5-7). Thus, a major question that arises is to what extent nuclear proximities between sites in the genome reflect specific long-range interactions with consequences for the transcriptional activity of underlying sequences.

The β-globin locus is flanked by two sites, HS5 and 3'HS1, that are occupied by the insulator protein CTCF. Chromosome conformation capture (3C) experiments have shown looping together of these sites in erythroid precursor cells before the globin genes are transcribed (8). Encompassed in this loop are the β-globin family of genes, 5' embryonic ε, fetal $^{A}\gamma$ and $^{G}\gamma$, and 3' adult β, and the locus control region (LCR) that activates them sequentially in erythroid cells (9). Globin genes are subsequently found in proximity to the LCR as they are activated during development (10). These looping interactions require erythroid transcription factors GATA-1, FOG-1 and EKLF, and the ubiquitous nuclear factor Ldb1, likely as participants in multimeric protein complexes bridging the LCR and genes (11-13). In addition, the chromatin remodeler Brg-1 is required at an early time in the looping process (14).

CTCF has been implicated in the organization of individual chromosomes and chromosome territories (15), and a role for CTCF in both organizational and transcriptional regulation has been suggested by the correlation of CTCF-mediated loops genome wide with lamin-associated domains demarcating regions of differing epigenetic marks (6). However, a functional role for CTCF site interactions has been probed in only limited instances (16). For example, the CTCF dependent imprinting control region (ICR) in the *Igf2/H19* locus forms an enhancer blocking loop with a site upstream of *Igf2* on maternal chromosomes, isolating the gene from enhancers it shares with H19 (17;18). Loss of CTCF or its sites results in loss of the loop and activation of *Igf2*. In contrast, individual deletion or disruption of β-globin HS5 or 3'HS1 CTCF sites did not affect β-globin transcription in mouse erythroid cells (19-22).

Recent integrative numerical frameworks have allowed the generation of spatial conformations of chromosomes *in silico* that are compatible with 3C-like high throughput data (23-28). More causal approaches have been used to study the mechanisms that may be responsible for chromosome organization by modeling chromatin as a polymer (7;29-34). In a theoretical analysis using such a polymer model, Mukhopadhyay *et al* suggested that long range interactions could have a crucial impact on the contact properties of genes and their enhancers (35). How chromatin configurations may influence gene expression remains unclear.

Here, we investigate whether regional long-range CTCF contacts contribute to transcriptional control at the β-globin locus. To this end, we constructed chromosome models that satisfy the constraints of long-range CTCF dependent interactions over 1.4 Mb of human chromosome 11 encompassing the β-globin locus (36). The models consist of a single, self-avoiding polymer chain along which specific sites (the CTCF sites here) are able to interact (30). Using chromosome conformation capture (3C) data as input, simulations of the best-fit models make it possible to quantify the spatial proximity between the globin genes and their LCR in erythroid K562 cells and non-erythroid 293T cells. The results show that the set of regional CTCF site interactions drive the LCR and globin genes closer together in expressing cells than in silent cells and nucleate a globule with the contacting LCR/globin gene pair deployed away from the center.

## MATERIALS AND METHODS

### Modeling the chromatin around the β globin locus: a self-avoiding worm-like chain with designated interacting sites

We simulated 1 Mbp long chromosome segments, each corresponding to the set of interacting CTCF sites (Figure 1), using a polymer model of the chromatin fiber (37) (Supplementary Figure 1). In this worm-like-chain (WLC) model, the chromatin fiber behaves as a flexible chain (37-39). We further excluded the possibility of the chromatin fiber overlapping itself (self-avoidance effect) by specifying the chain diameter to be 30 nm. The base-pair density along the fiber was fixed at 150 bp/nm (37). In order to prevent the formation of knots, simulated chromosomes were circularized by adding a 0.2 Mbp long fiber free of CTCF sites (not represented in the figures), which is long enough (1.3 μm) to prevent looping of the ends of the 1 Mbp chromosome. This led to a final 1.2 Mbp chromosome segment, which was confined to a spherical volume of 1 μm diameter.

Along the chromatin, designated sites akin to CTCF sites interact with each other according to a harmonic potential $V(x) = \frac{k}{2}x^2$ whose value is lowest for a spatial distance between the sites, *x*, that is equal to zero. For any given pair of known interacting sites, the spatial distance, *x*, separating them varies according to the chromosome conformation and the larger the intensity, *k*, of their potential, the more effective the potential *V(x)*, that is, the more frequently the sites interact. Unlike models where interactions are short-range (30), long-range potentials such as harmonic ones prevent interacting sites from being spatially distant from each other. This allowed us to comprehensively study the statistical properties of the models by running numerous short (*i.e.*, < 1 day) simulations.

**Fitting the polymer models: an iterative procedure**

3C experiments quantify interaction frequency between two sites. For input to the models, these values were converted to potential intensities. To this end, the potential intensities for a given set of 3C contact frequencies were iteratively updated to obtain contact frequencies *in silico* as close as possible to contact frequencies *in vivo* (see Supplementary Material). In addition, we do not consider interactions between sites that are in contact with a common third site in order to avoid non-linear effects on the uniqueness of the solution of the iterative algorithm (Supplementary Figure 2).

**Computing statistical properties of best-fit polymer models**

Given a best-fit polymer model, we used distributed computer facilities in order to sample an equilibrium set of conformations as large as possible. An equivalent computation time of 3100 days (~ 8.5 years) was run on the European Grid Infrastructure (EGI) using two virtual organizations: (VO) biomed (biomedical computing) and vo.iscpif.fr (complex systems computing, see http://www.iscpif.fr/tiki-index.php?page=csgrid). The use of the distributed facilities was accomplished through the OpenMOLE project (http://www.openmole.org/) (40).

For a more detailed description of the methods see Supplementary Information.

## RESULTS

### CTCF site long-range interactions surrounding the β-globin locus are primarily cell type specific

In previous work, we used chromatin immunoprecipitation combined with microarray analysis (ChIP-chip) and ChIP and qPCR to survey and validate CTCF sites up- and downstream of the β-globin locus on human chromosome 11 in K562 cells, where the fetal γ-globin gene is transcribed and in 293T cells where the globin locus is silent (36). A wealth of structural information is available for K562 cells through the ENCODE consortium and these cells demonstrate the appropriate LCR-γ-globin chromatin looping while 293T cells do not (41)(C.M. Kiefer and A. Dean, unpublished). All sites analyzed were similarly occupied by CTCF in the two cell types and all were co-occupied by cohesin complex members Rad21, SMC1 and SMC3. We then used 3C to determine, in these two cell types and in K562 cells in which CTCF had been reduced using RNAi (K562kd), the frequencies of interaction among all occupied CTCF sites surrounding the globin locus over a chromosomal range of 1.4 Mbp. 3C quantifies the interaction frequency between two sites in chromatin after formaldehyde cross-linking of protein and DNA, restriction enzyme digestion, re-ligation and detection of novel sequence junctions by qPCR (42).

A representation of the interactions previously observed (36) is presented in Figure 1. Each black dot represents an occupied CTCF site. Each parabola represents a pairwise CTCF site interaction where the height of the curve indicates the relative frequency of interaction on a log scale. Also represented are the decreased interactions between these sites when CTCF is knocked down in K562 cells by RNAi, reducing γ-globin transcription and allowing incursion into the locus of the repressing histone modification H3k9me2. Although CTCF was bound at nearly identical locations throughout the locus in the different cell types, cell type specific patterns of long-range interactions distinguish K562 cells (red) from 293T cells (blue) with 60% of interactions being cell type specific. Interestingly, the loop between β-globin 3'HS1 and HS5 (C15 and C16 respectively) was observed in both cell types. In K562kd cells, all interaction frequencies were significantly reduced compared to CTCF-replete cells but the pattern of interactions was unchanged. These data raise the possibility that CTCF organizes the chromosomal domain structure surrounding the globin locus to influence transcription.

**Modeling cell type specific interactions between CTCF sites**

To address the issue of whether differential cell type specific CTCF contacts affect the conformation of the globin region, a polymer model of the 30 nm chromatin fiber was designed for each of the three cell types under investigation: K562, K562kd and 293T. Disregarding its internal structure, chromatin behaves as a flexible chain that self-avoids. We used the simplest continuous model, known as the worm-like chain (WLC), which includes flexibility properties of a single chain (37-39) and considers the self-avoiding nature of a 30 nm thick fiber. Each WLC model represents a 1Mbp region of human chromosome 11 containing the interacting CTCF sites and the β-globin locus. Interaction forces were determined by iterating the DNA folding algorithm until the frequency of contacts between the sites corresponded to the experimentally measured CTCF dependent chromatin contacts (Figure 1) (36). Importantly, CTCF interaction frequencies were the sole input to the polymer models and the inputs did not include the LCR or any gene promoter sites. This allowed us to make independent *in silico* measurements of LCR-promoter proximity during the model runs using these proximities as a “read-out” of the potential transcriptional state of the locus.

3C experiments measure interactions in a population of cells. We assume that the measured population-averaged frequencies are analogous to time-averaged frequencies at equilibrium in the models – that is, averaging over populations is equivalent to averaging over time. Two parameters of the model are adjustable, namely, the strength of interaction between the CTCF sites (interaction potential) and the persistence length, which describes the flexibility status of the chromatin fiber—the larger the persistence length, the less flexible the chromatin fiber (Supplementary Figure 1). Since the persistence length is not precisely known *in vivo*, we studied three values (100 nm, 200 nm and 300 nm) typical of those reported in the literature (38). We then iteratively and independently adjusted intensities of the interaction potentials so that equilibrium contact frequencies between CTCF sites best matched experimentally measured frequencies (36). These best-fit interaction potentials were subsequently used in a benchmark polymer model to compute statistical properties such as equilibrium distributions of the spatial distances between the globin genes and the LCR and typical chromatin conformations.

The pattern of intensities obtained for the interaction potentials in K562 and K562kd cells were comparable (Supplementary Table 1), with intensities reduced in K562kd cells compared to K562 cells. This indicates that interactions are based on a similar but weakened mechanism, in accord with the incomplete knockdown by RNAi which allows some CTCF protein molecules to remain associated with chromatin (36). By contrast, intensities obtained for the potentials in 293T cells were globally much lower than those in K562 cells, consistent with the overall lower frequency of 3C interactions observed in 293T cells. More specifically, two categories of interactions of qualitatively different natures can be distinguished in 293T cells. The first category consists of the pairs C-08/C-20 and C-20/C-21 (see Figure 1) for which the interaction intensities were relatively strong. The second category contains all the other pairs of interacting sites for which the intensities were at least two orders of magnitude lower (Supplementary Table 1).

**Sites within the globin locus domain are in closer proximity to the LCR in erythroid cells than in non-erythroid cells**

Next, we sought to tie together the modeled interactions in these cells with known expression patterns. Since the primary difference between the cell types is the expression of globin genes, and interaction with the LCR is required for transcribing these genes, we chose to use the proximity of the LCR to a series of targets in the locus as a proxy for transcriptional activity. The primary set of targets consisted of the transcription start site (TSS) of each of the globin genes in the locus. In addition, we included the TSS for the β-globin pseudogene, two intergenic sites, and a long-distance site downstream from β-globin (Figure 1).

The distributions of 3-D spatial distances between the LCR and each gene were measured for each cell type over the course of 100 simulations of the corresponding best-fit model. Frequencies and cumulative frequencies show the time fraction two sites spend at a given distance or at a distance lower than a given distance, respectively. These are displayed in Figure 2 for a representative target, the $^{G}\gamma$-globin gene, for all simulations (thin curves) and for the corresponding averages (thick curves). Figure 2A demonstrates a clear shift of the distributions toward large distances for 293T cells compared to K562 cells. Looking across all targets (Supplementary Figure 3), as genomic distance from the LCR increases, so does the 3-D spatial distance. This effect is stronger in 293T cells than in K562 cells so that at the largest genomic distances there is the greatest difference between cell types. Cumulative frequencies, shown on a log scale, reveal a tendency for the globin genes to be closer to the LCR in K562 than in K562kd (Figure 2B, Supplementary Figure 3). Importantly, the results do not qualitatively depend on microscopic parameters (the persistence length) of the chromatin fiber. Taken together, these results show that in K562 cells, regional CTCF sites interact in such a way that globin genes are closer to the LCR than in 293T cells, and to a lesser extent, than in K562kd cells.

**Effects of chromatin features and genomic organization on LCR contacts**

We next studied how the physical features of the chromatin fiber affect the capacity of globin genes to make contact with the LCR. We compared results obtained for three levels of stiffness of the chromatin fiber typical of those reported *in vivo,* which are respectively defined by persistence lengths equal to 100 nm, 200 nm and 300 nm (Figure 3). Since sites are likely to interact through protein complexes (*e.g*., transcription factors, RNA polymerase), we chose a value of 10 nm to represent the space taken up by these complexes, and defined an interaction event to exist when the LCR was within 10 nm of a target site, *i.e*., when there was <=40 nm between the centers of the two 30-nm chromatin fibers. We then quantified the frequency of LCR-target interaction events across the locus.

The frequency of an interaction event can be shown on a negative log scale so that higher values indicate greater isolation of the LCR from targets. This analysis highlights two important points (Figure 3). First, the stiffer the chromatin fiber, the larger the difference between 293T cell and K562 cell contact frequencies. For example, the isolation of $^{G}\gamma$ from the LCR is ~ 20-fold larger in 293T than in K562 at a persistence length of 100 nm; it is at least 1000-fold larger at a persistence length of 300 nm, which is within the range of the expected persistence lengths *in vivo* (38). Second, the smaller the genomic distance to the LCR, the larger the difference in contact frequencies between K562 and K562kd cells. For example, at a persistence length equal to 200 nm, LCR contacts with β (genomic distance = 53 kbp) are only ~1.5-fold less frequent in K562kd cells than in K562 cells; contacts with γ (genomic distance = 26 kbp) are ~ 4.5-fold less frequent. The observation that the $^{G}\gamma$-globin gene is more isolated from the LCR in K562kd cells than in CTCF-replete K562 cells is consistent with the reduction of $^{G}\gamma$-globin expression by RNAi mediated CTCF knock down (36).

**Spatial conformations of the globin locus during active transcription of the γ-globin genes**

Compared to 293T cells, CTCF interactions in K562 cells provide an overall modeled conformation of the globin locus that favors contact between the LCR and globin genes (Figure 3). We ran 100 additional simulations of a K562 polymer model to study the topology of the chromatin conformations that involved LCR-globin gene interactions. For each simulation, snapshots were stored of the first configuration that was reached when the LCR was in contact with a globin gene of interest. Figure 4A shows one such configuration for $^{G}\gamma$ (see Supplementary Figure 4 for other genes).

Like the 3C contact maps to which interaction potentials were fitted (Figure 1), the snapshots reveal conformations that are composed of one globule, where many contacts are present (blue sites), plus one loop where only two sites interact (red sites). These two structures tend to repel each other, due to volume exclusion effects (43). Interestingly, the contacts between the LCR and $^{G}\gamma$ gene (green sites) occur most often on the periphery of the globule. We quantified this with a radial distribution function, which shows the probability that a site is found at a certain distance away from the center of mass of the globule.

Whereas the LCR does not have, on average, a preferred location with respect to the center of mass of the globule (compare red and black dotted curves in Figure 4B), interaction events between $^{G}\gamma$ and the LCR tend to happen farther out from the center of the globule than is typical for the locus as a whole (Figure 4B, blue; note that the blue line is less smooth since LCR-$^{G}\gamma$ interactions only occur during a subset of all time steps represented by the red and black curves). Extending this analysis to $^{A}\gamma$ and β indicates that all globin genes, but particularly γ-globin genes, tend to be located more peripherally to the globule regardless of LCR contact (Figure 4C). By contrast, in 293T cells, where the globule is less compact, no preferential location is observed for any of the locus sites of interest (Figure 4D). These findings suggest that, in addition to favoring contacts with the LCR, the CTCF-driven globule in K562 cells tends to displace the genes to be activated, *i.e.*, the γ-globin genes here, away from the surrounding chromatin.

**Dominant CTCF interactions and stiff chromatin prevent contacts between the LCR and globin genes in 293T cells**

The interaction potentials observed in 293T cells can be divided into two categories based on strength (Supplementary Table 1). The strongest potentials are between C-08 and C-20 and between C-20 and C-21. A polymer model where these interactions alone are present leads to a reduction of the tendency for globin genes to be spatially close to the LCR when the chromatin fiber is stiff (Supplementary Figure 5). To investigate the influence of these interactions, in particular whether the strongest interactions found in 293T cells are sufficient to decrease LCR-gene interactions compared to K562 cells, we used two additional models: one where only the two strongly-interacting sites are present (ignoring all other interactions measured by 3C in 293T cells) and another using chromatin with no interacting sites. Since the interaction events we defined earlier (40 nm between chromatin fiber centers) do not always occur in 293T cells as they do in K562 cells, we used the minimal distance obtained in 100 simulations as an alternative metric to represent LCR-target proximities.

The model with no interacting sites serves as a baseline (red lines, Figure 5). One might hypothesize that introducing any interacting sites in this locus would bring the LCR closer to targets on average. However, interestingly, the model with just two pairs of strongly interacting sites results in a substantial isolation of the LCR from the targets over most of the locus compared to the model with no interacting sites (compare purple and red lines, Figure 5). This strong isolating effect decreases when the rest of the 293T interaction potentials are included (blue lines, Figure 5), even though the short minimal distances reported here remain very infrequent (see Figure 3). These additional interactions tend to re-establish contacts of a subset of the globin genes, *e.g.*, β (Figure 5, upper panel). The isolating effect of the two pairs of strongly interacting sites suggests that they may play an important role in cell-type specific chromosome architecture surrounding the β-globin locus.

## DISCUSSION

Above the megabase scale, Hi-C studies have revealed that contact maps of individual human chromosomes can be divided into two or three classes of regions, with one class that is more transcriptionally active than the other(s) (7;44). Novel conformation capture techniques have further revealed that inter-chromosomal contacts not only correlate with transcriptional activity but also with CTCF-bound regions (26;44). Between the megabase scale and the nucleosome level of chromatin folding little is known about chromatin organization. In this study, we use a thermodynamic framework based on a worm-like chain model of chromatin where designated sites along the DNA, here CTCF sites over 1 Mbp around the β-globin locus, are able to form pairwise interactions (30;36). The results indicate that (i) proximity within the globin locus between the LCR and genes and (ii) displacement of the active LCR/globin pair away from surrounding chromatin are both facilitated by overall erythroid cell type specific CTCF chromosome organization. Thus, providing a thermodynamic topological model with experimental measurements of interaction frequencies of CTCF sites alone predicts a chromosome conformation pattern consistent with the known biology of the human β-globin locus.

Our analysis uses long-range interaction potentials to efficiently sample conformations that are consistent with the CTCF contact map. The main advantage of this choice is that it allows a comprehensive statistical analysis of chromosome folding. The disadvantage is that real interactions have a shorter range in the cell, and the choice of long-range interactions may alter the statistical properties of the simulated chromosome conformations. Nevertheless, long-range interactions might mimic the real situation in cells, where non-functional chromosome conformations appear to be dis-favored by mechanisms that remain elusive.

CTCF sites within 1 Mbp of the β-globin locus were occupied in both K562 (active locus) and 293T (silent locus) cells, however, their pair-wise interactions were primarily cell type specific, suggesting that factors other CTCF are important for the specificity of loop formation. In some loci, cohesin co-occupancy is the determinant of CTCF insulator function, however, cohesin was present the at CTCF sites we studied in both K562 and 293T cells (45). Modifications to these proteins or recruitment of additional components may explain the difference in loop formation between the cell types. In addition, looping interactions beyond those we studied are likely to impinge on chromosome organization in the β-globin region, including those mediated by different protein factors (46;47). Nonetheless, our simulations indicate that consideration of CTCF looping contacts alone predicts a cell type specific chromosome organization in which the probability of intra-loop LCR/globin gene contacts is favored for K562 cells and hindered in 293T cells. Interestingly, numerous interacting sites around the α-globin locus were found to be bound by CTCF protein and our previous data revealed that α-globin transcription was reduced upon CTCF knock down (36;48). This further reinforces the idea of an important role of regional CTCF sites in the interplay between the architecture of chromosomes and their function.

The modeled distances from the LCR to globin genes as a whole are smaller in K562 cells, where the locus is active, than in 293T cells where the locus is inactive. Moreover, reduction of CTCF in K562 cells using RNAi, which leads to a reduction of γ-globin transcription (36), increases these modeled distances compared to cells replete with CTCF. Thus, CTCF site contacts in K562 cells may provide a constraint such that the collision dynamics between globin genes and the LCR are the most appropriate for LCR-globin gene interactions. We suggest that the relative closeness of the LCR to the coding regions of the locus modeled in K562 cells, compared to 293T cells, favors 'sampling' of contacts between the LCR and genes that, if stabilized by protein-protein interactions, allow looping and transcription activation. Our results indicate gains in sampling time that can exceed 1000-fold, depending on the stiffness of the chromatin and the position of the gene. In this view, the active K562 γ genes would successfully establish LCR contact, whereas, the silent β gene would not, due to the modulation of required developmental stage specific proteins or protein complexes. The data do not exclude the possibility that LCR/globin gene contact is the initial organizing event for the locus during erythroid differentiation and that the CTCF looping ensemble evolves to stabilize this interaction. Reduction of the LCR/globin gene interaction when CTCF is decreased by RNAi shows the importance of the CTCF-mediated loop ensemble to maintain contact.

For transcription to occur, the LCR and gene to be activated must be accessible to the transcription machinery. Our studies highlight in K562 cells a modeled preferential exposure of the LCR and γ genes to the nuclear milieu at the surface of a CTCF-driven globule. This positioning is reminiscent of extrusion from a chromosome territory and might facilitate localization of the γ-globin/LCR ensemble within nuclear foci of transcription to which globin genes are known to migrate to achieve high levels of transcription (49-51). Interestingly, modeling of chromosome conformation capture carbon copy (5C) interactions in K562 cells suggested that the α-globin genes cluster together with neighboring active housekeeping genes that surround them in the interior of a globule that might correspond to a transcription factory (48). By contrast to the α-globin locus, the β-globin locus is embedded among odorant receptor genes that are silent in erythroid cells. Thus, the tendency for the active γ-globin gene and LCR to be on the periphery of a globule might be related to the difference in transcriptional state compared to neighboring genes.

This work highlights the interplay of numerous CTCF sites within a chromosomal region that interact to form multiple chromatin loops. The modeled conformation of the region reveals an ensemble of interactions (the globule), whose overall distribution may have more subtle biological implications (e.g., the propensity for $^{G}\gamma$ to be on the periphery) than suggested by consideration of individual loops. This may help to explain why individual deletion or disruption of HS5 or 3'HS1 (C-16 and C-15, respectively, in Figure1) did not affect β-globin transcription in mouse erythroid cells (19-22). Likewise, in this view, the insertion of an extra HS5 site (52) that creates two new loops and limits access of the LCR to globin genes could be interpreted as introducing an imbalance to the ensemble of interactions, similar to the two pairs of strong interacting CTCF sites in 293T cells that isolate the LCR from the locus. These ideas suggest specific, testable biological hypotheses. For example, disrupting the strongly interacting CTCF site pairs in 293T cells should reduce LCR isolation from globin genes, although such manipulation is unlikely to affect transcriptional activity due to the absence of erythroid transcription activators in these cells. Alternatively, incremental deletion of CTCF sites in the K562 computational model could suggest which sites are most important for LCR/globin gene contact, and the predictions could then be tested experimentally.

## SUPPLEMENTARY MATERIAL

Supplementary Data are available at NAR online: Supplementary Table 1, Supplementary Figures 1-7, and Supplementary Methods.

## FUNDING

This research was supported by the Intramural Program of the National Institute of Diabetes and Digestive and Kidney Diseases, National Institutes of Health (KIA 15508 to AD). I. J. is supported by a Novartis grant (CRG) and thanks Région Île-de-France and ISC-PIF for financial and logistic support. This work was also supported by the Sixth European Research Framework (project number 034952, GENNETEC project), PRES UniverSud Paris, CNRS and Genopole (F.K.).

## ACKNOWLEDGEMENTS

We thank Olivier Martin and Elissa Lei for helpful comments on the manuscript.

## FIGURE LEGENDS

**Figure 1.** CTCF based chromatin loops differ between K562 cells and 293T cells. CTCF site interactions from Hou et al. 2010 (36) are depicted in a schematic diagram of the globin locus. Pairwise interaction frequencies detected by 3C are shown by arcs for K562 (red), K562 CTCF RNAi knockdown (green) and 293T (blue) cells. Black dots below the x axis indicate CTCF-occupied sites confirmed by qPCR used in this analysis. The LCR is indicated by a gray vertical line. RNA seq data from the ENCODE Consortium (CalTech data set from Wold and Myers groups) for K562 cells (red track) and 9 other non-erythroid cell types combined (multi-color track) are shown as normalized read density (reads per million, RPM) on a ln(x+1) scale. RefSeq gene models for the locus are shown, including globin genes (orange), odorant receptor genes (blue) and other genes (gray). All coordinates and data are for the hg19 assembly.

**Figure 2.** Distribution of distances between $^{G}\gamma$ and the LCR using a best-fit polymer model of the β-globin locus. The curves show the distributions of the spatial distances (in nm) separating the LCR from $^{G}\gamma$ in the three cell types K562, K562kd and 293T, for a chromatin fiber with a 200 nm persistence length. Thin dotted curves are distributions obtained from 100 different runs, which were used to compute the average curves (thick solid curves). (A) The y-axis reports the time fraction the two sites spend at the distance reported along the x-axis. (B) The y-axis reports the time fraction the two sites spend at a distance lower than that reported along the x-axis. Note that at distances lower than 100 nm, the cumulative frequency is 10 to 100 times larger in K562 cells than in 293T cells.

**Figure 3.** Interaction frequency as a function of the genomic distance to the LCR. The panels show the co-logarithm of the contact frequency between the LCR and β-globin genes, the intergenic sites (i1, i2) and the downstream site (ds) for three persistence lengths. Higher values indicate increased isolation of LCR from target. A contact is defined whenever the distance between the sites along the polymer and the LCR is lower than 40 nm. This figure shows that the stiffer the chromatin, the lower the contact frequency in 293T cells. Note that due to numerical limitations, frequencies lower than $10^{-6}$ cannot be detected: they are reported to be equal to $10^{-6}$ (*e.g.*, for $^{G}\gamma$ at a persistence length of 300nm)

**Figure 4.** Chromatin conformations favoring contacts between the β-globin genes and LCR in K562 cells. (A) Typical conformation of the 1Mbp region around the β-globin locus during a contact between LCR (green+star) and $^{G}\gamma$ (green). Blue sites: CTCF sites that form a connected network of interaction (Supplementary Figure 1). Darkest blue sites: CTCF sites that surround the β-globin locus. Red sites: the isolated interaction between C-08 and C-10. The conformation can be divided into a loop (stabilized by the red sites) and a compact globule (dashed orange ellipse) encompassing the region from C-03 to C-10. (B) Spatial location of the contact: using 1000 equilibrium simulations of the same best-fit polymer as in (A), we report i) the radial mass distribution of the compact globule, *i.e.* the average probability density for the location of the C-03 to C-10 region with respect to the globule center of mass; ii) the radial distribution of $^{G}\gamma$ and LCR during contacts and iii) the radial distribution of the LCR (no matter the position of $^{G}\gamma$). One can see that the $^{G}\gamma$/LCR contacts tend to occur away from the globule center. (C) Spatial location of the globin genes in K562 (obtained from 100 simulations of the best-fit polymer). Genes tend to be located away from the center regardless of LCR contact. Large distances are particularly enhanced in the case of the γ genes. (D) Same as in (C) but for 293T cells. No particular location can be observed for any of the genes.

**Figure 5**. Minimal distances between the LCR and the β-globin genes. Each point represents the minimal spatial distance obtained in our simulations between a globin gene and the LCR at persistence lengths 200nm and 300nm. In K562 cells, the minimal distance is 30 nm, indicating that physical contacts are always possible. The situation is different for 293T cells. In particular, when only the two strongest potential pairs in 293T cells are considered (“2pairs”), the minimal distance can be as large as 80 nm (purple points). The red points show the results where no interaction is considered. By taking into account all the interactions involved in 293T cells, one can observe for persistence lengths of the chromatin around 300 nm a complete isolation specific to $^{G}\gamma$ and points linearly closer to the LCR.

## Figure 1

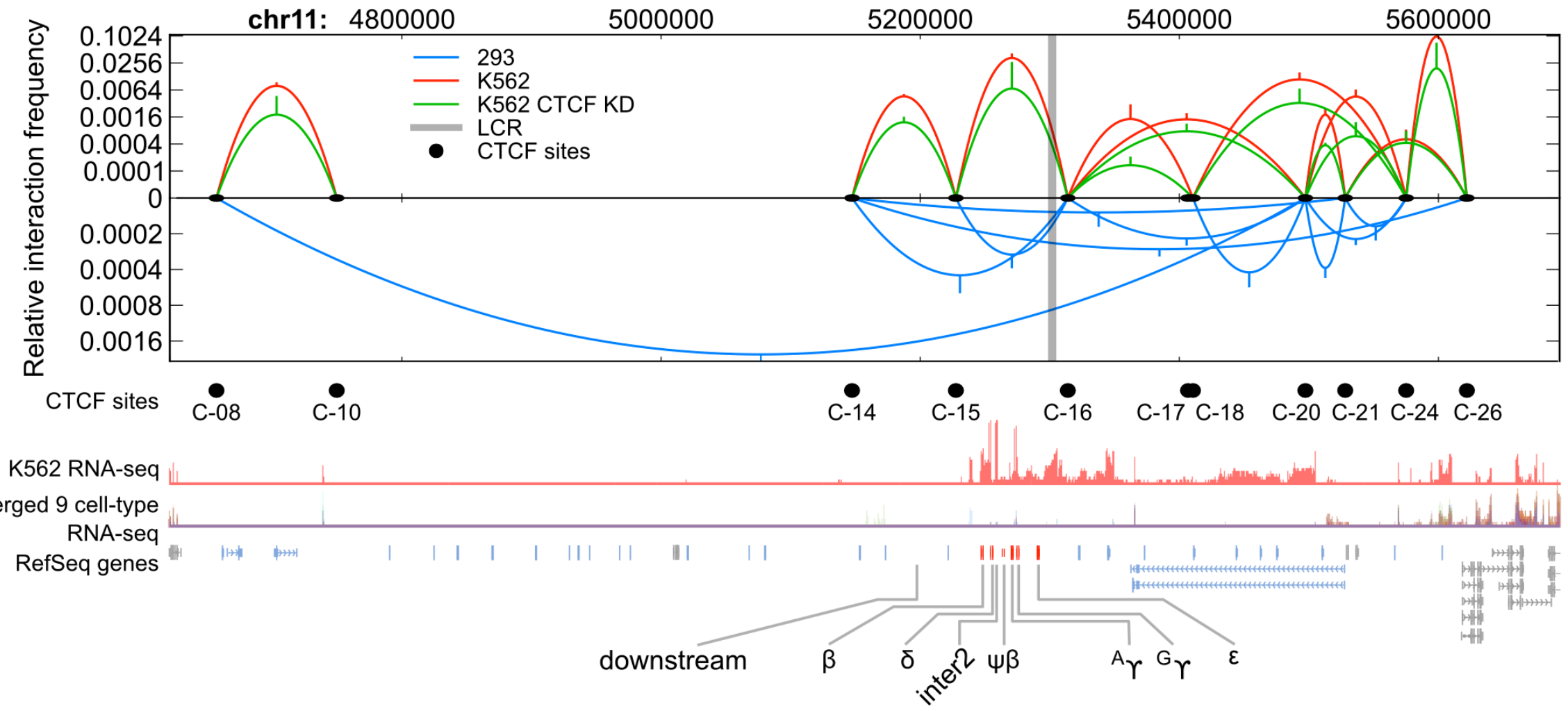

**Figure 2**

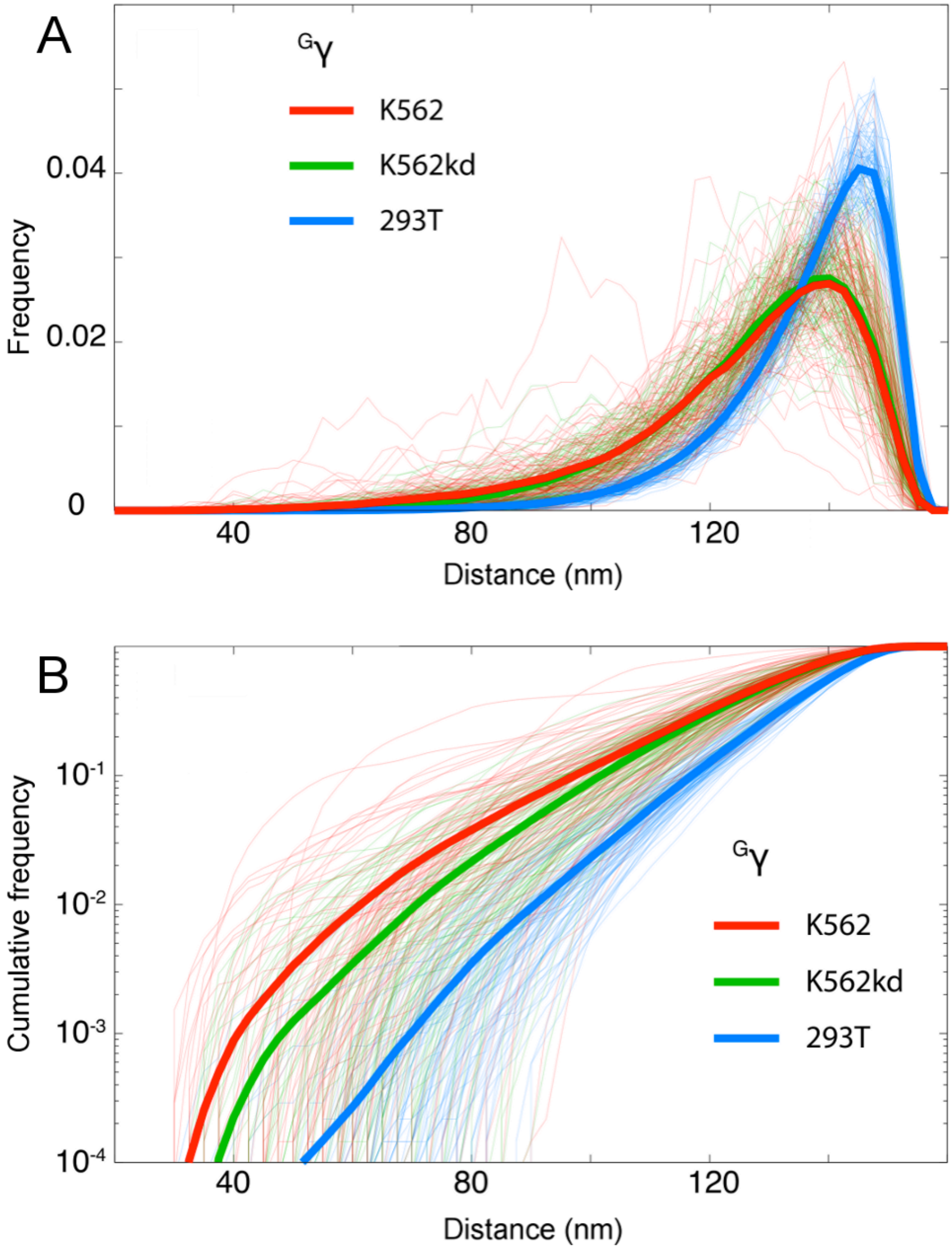

**Figure 3**

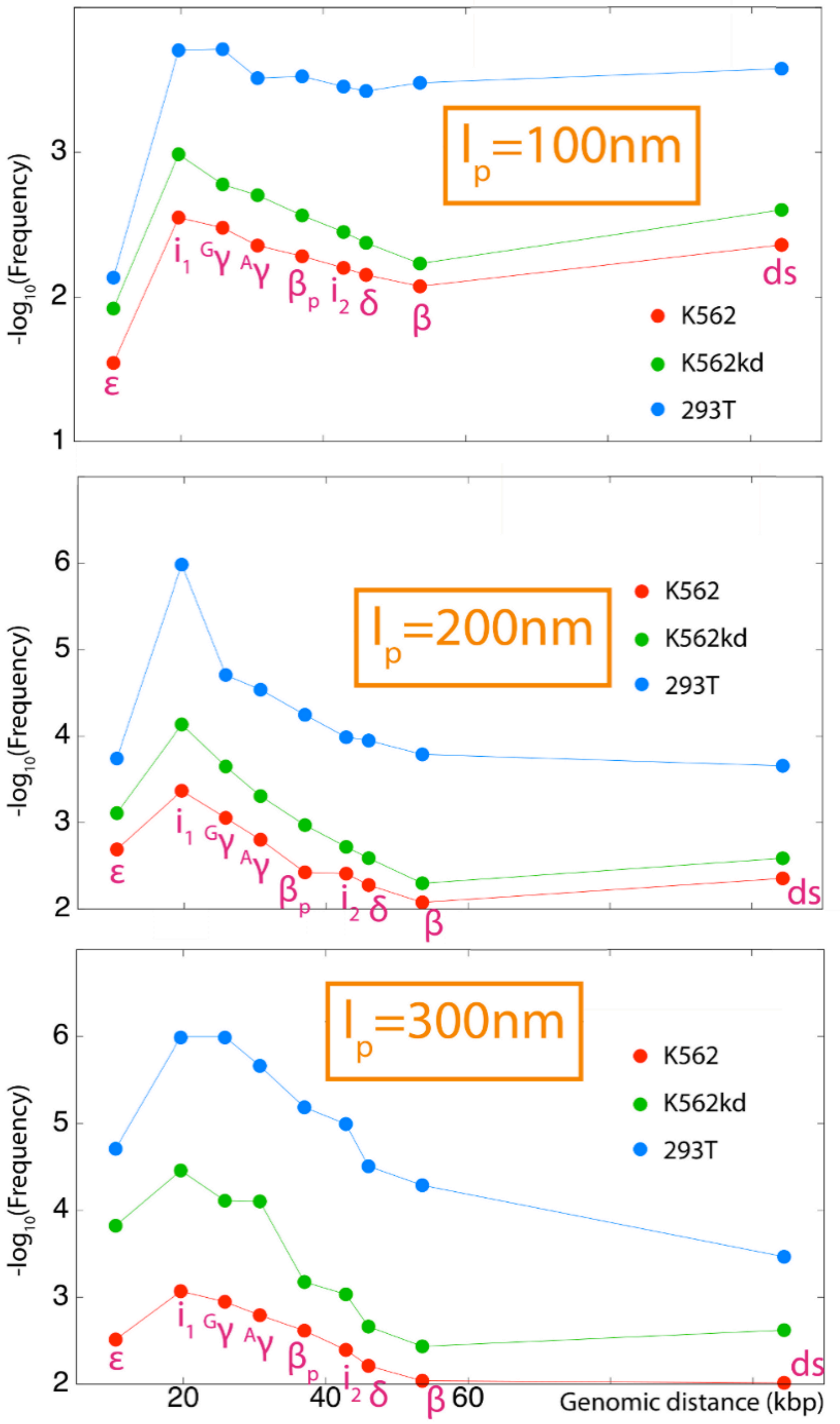

**Figure 4**

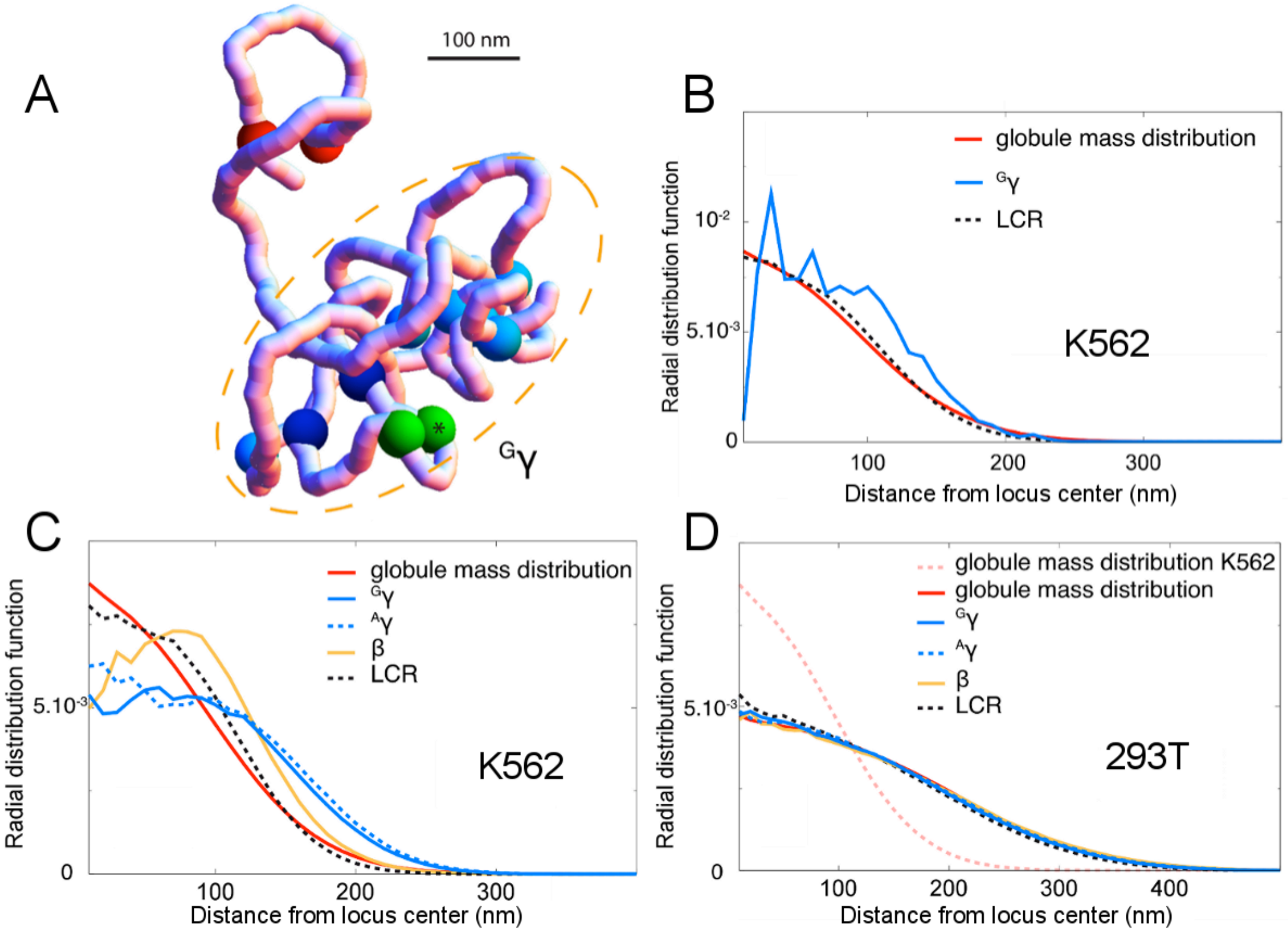

**Figure 5**

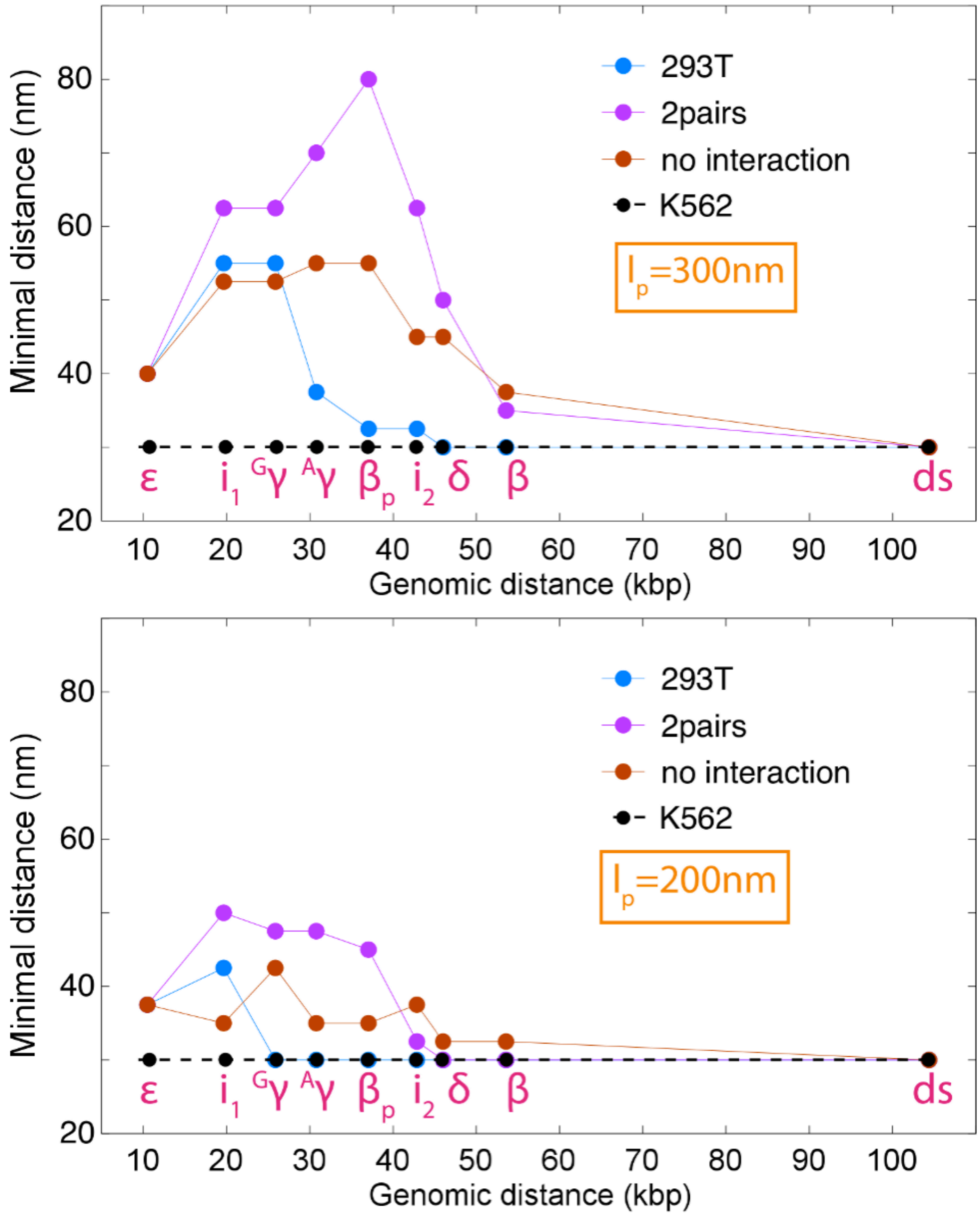